\documentclass[aip,jcp,reprint,superscriptaddress]{revtex4-1}

\usepackage{amsmath} 
\usepackage{graphicx} 
\usepackage{float}
\usepackage{helvet}

\renewcommand{\vec}{\bf}

\begin{document}

\title{Combining phonon accuracy with high transferability in Gaussian approximation potential models}

\author{Janine George}
\affiliation{Institute of Condensed Matter and Nanosciences, Universit\'e catholique de Louvain, Chemin des \'Etoiles 8, 1348 Louvain-la-Neuve, Belgium}

\author{Geoffroy Hautier}
\affiliation{Institute of Condensed Matter and Nanosciences, Universit\'e catholique de Louvain, Chemin des \'Etoiles 8, 1348 Louvain-la-Neuve, Belgium}

\author{Albert P. Bart\'ok}
\affiliation{Department of Physics and Warwick Centre for Predictive Modelling, School of Engineering, University of Warwick, Coventry CV4 7AL, United Kingdom}

\author{G\'abor Cs\'anyi}
\affiliation{Engineering Laboratory, University of Cambridge, Cambridge CB2 1PZ, United Kingdom}

\author{Volker L. Deringer}
\email{volker.deringer@chem.ox.ac.uk}
\affiliation{Department of Chemistry, Inorganic Chemistry Laboratory, University of Oxford, Oxford OX1 3QR, United Kingdom}

\begin{abstract}
    Machine learning driven interatomic potentials, including Gaussian approximation potential (GAP) models, are emerging tools for atomistic simulations. Here, we address the methodological question of how one can fit GAP models that accurately predict vibrational properties in specific regions of configuration space, whilst retaining flexibility and transferability to others. We use an adaptive regularization of the GAP fit that scales with the absolute force magnitude on any given atom, thereby exploring the Bayesian interpretation of GAP regularization as an ``expected error'', and its impact on the prediction of physical properties for a material of interest. The approach enables excellent predictions of phonon modes (to within 0.1--0.2 THz) for structurally diverse silicon allotropes, and it can be coupled with existing fitting databases for high transferability. These findings and workflows are expected to be useful for GAP-driven materials modeling more generally.
\end{abstract}

\maketitle

\section{Introduction}

Vibrational properties on the atomic scale determine the thermal behavior of materials. Their knowledge is therefore of central importance in many fields of physics, materials science, and engineering. For example, a low thermal conductivity is a requirement for thermoelectric waste-heat recovery, \cite{Snyder2008} and the ability to predict this property based on highly accurate theoretical and computational methods can allow the community to discover possible new thermoelectric materials. Indeed, thermal conductivity and other macroscopic quantities can nowadays be computed from first principles, normally based on density-functional theory (DFT), although at a very substantial computational cost. \cite{Carrier2007, Stoffel2010, Hellman2013, Skelton2014, Togo2015, Mizokami2018} Only very recently, larger DFT-computed databases of harmonic vibrational properties became available, \cite{petretto2018high, togo} but high-throughput predictions of more computationally expensive properties such as thermal conductivities have not been attempted to our knowledge. So far, computational searches for materials with high thermal conductivity have been performed based on computationally cheaper models such as the quasi-harmonic Debye model, that do not require the full \textit{ab initio} computation of the thermal conductivity, or combined with global optimization techniques, that only require a limited amount of full \textit{ab initio} computations of the thermal conductivity.\cite{Toher2014,Seko2015a}

In an effort to sidestep the computational cost of DFT, machine learning (ML) based interatomic potential models are increasingly widely used in materials modelling. \cite{Behler2017, MLP_AdvMater, Zuo2020, Mueller2020} Using a reference database of (typically) DFT data and a regression framework including artificial neural networks,\cite{Behler2007, Artrith2012, Smith2017, Zhang2018a} kernel methods, \cite{Bartok2010, Huan2017, Chmiela2017} or linear fitting, \cite{Thompson2015, Seko2015, Shapeev2016} they enable atomistic simulations at similar accuracy levels but at orders of magnitude lower computational cost. The high accuracy that ML potentials can reach for phonons was demonstrated a decade ago already, \cite{Eshet2010} and early applications to amorphous phases showcased the ability to treat large and structurally complex systems. \cite{Sosso2012a, Sosso2018} To date, vibrational properties continue to be a sensitive test for the quality of a candidate potential, \cite{Rowe2018, Bartok2018, Zhang2019, Marques2019} because they give a direct and physically meaningful measure for how reliably the interatomic forces are predicted by any ML model.

Beyond the prediction of harmonic phonons in crystals, ML potentials have begun to be used for thermal properties including anharmonic effects as well. A proof-of-concept in the Gaussian Approximation Potential (GAP) framework, for the temperature-dependent phonon dispersion curves of Zr, was reported in 2018. \cite{Qian2018} Skutterudite CoSb$_3$ (Ref.\ \citenum{Korotaev2019}) and elemental metals \cite{Ladygin2020} were studied using Moment Tensor Potentials (MTP) very recently, showing excellent agreement with DFT reference values. Three independent studies dealt explicitly with crystalline diamond-type ({\bf dia}) silicon. \cite{Babaei2019, Minamitani2019, Qian2019}

An aspect that so far has been rarely discussed in this context is the ability of materials to crystallize in different structures (``allotropes'' for elements, ``polymorphs'' for compounds). The evaluation of phonon properties of silicon with ML potentials has focused, so far, exclusively on the most abundant structure, {\em viz.} the {\bf dia} allotrope. \cite{Babaei2019, Minamitani2019, Qian2019} There is, however, ample interest in others: clathrate-type structures have been extensively studied by DFT; \cite{Tang2006, Karttunen2011, Karttunen2013} many other possible silicon allotropes have been proposed. \cite{Zhao2013, Mujica2015, Amsler2015, Saleev2017, Jantke2017, He2018} An open-framework structure, oS24, was synthesized by de-intercalation from Na$_4$Si$_{24}$; \cite{Kim2015} it was later studied specifically with respect to its thermal properties, using DFT. \cite{Ouyang2017} Other metastable silicon allotropes have been observed in laser-induced transformations. \cite{Rapp2015} An overview, including possible synthesis routes, was given recently. \cite{Haberl2016}

In the present work, we explore how phonon properties for a diverse ensemble of crystal structures (allotropes or polymorphs) can be described in the GAP framework. We discuss computational protocols by which reference databases can be assembled, and potentials fitted, to deal with a range of silicon allotropes. Our work outlines a general strategy for generating GAP models that can be interfaced to high-throughput materials workflows.

\section{Methodology}\label{sec:methods}

\subsection{Gaussian Approximation Potentials} 

Interatomic forces, on which all of the present study is based, were obtained in the GAP framework. \cite{Bartok2010} Initially, we tested the potential model for silicon developed by Bart\'ok et al. (Ref.\ \citenum{Bartok2018}), which has been designed as a ``general-purpose'' interatomic potential for various applications in physics and materials science, and which has been extensively validated for physical properties of crystalline \cite{Bartok2018} and amorphous \cite{aSi_structures_GAP, Bernstein2019} silicon. We refer to this potential as ``GAP-18'' in the following. In the present work, we describe two methodological advances over that previous study. First, we developed new fitting databases by various strategies, with a specific view to describe vibrational properties, as detailed in the Results and Discussion section. Second, we use an atom-wise adaptive regularization scheme to improve the accuracy of the fit.

In brief, and using the notation of Ref. \citenum{Bartok2018}, the energy in GAP-18 is fitted as
\begin{equation}
    E = \sum_{i<j} V^{(2)}(r_{ij}) + \sum_{i} \varepsilon_{i}, 
    \label{eq:E}
\end{equation}
where the first sum is a baseline pair potential to capture exchange repulsion at short interatomic distances, and the second sum is given by the Gaussian process regression itself. The atomic energy, $\varepsilon_{i}$, of a given atom, unknown from DFT but the key quantity in GAP, is
\begin{equation}
    \varepsilon_{i} = \sum_{s}^{M} \alpha_{s} K\left( \mathcal{R}_{i}, \mathcal{R}_{s} \right), 
    \label{eq:epsilon}
\end{equation}
based on a kernel (or similarity) function, $K$. The latter compares the local environment of the $i$-th atom in a given structure, $\mathcal{R}_{i}$, with all $M$ environments in a sparse reference set, $\mathcal{R}_{s}$. In GAP-18, this kernel function is given by the Smooth Overlap of Atomic Positions (SOAP) formalism; \cite{Bartok2013} we use the same SOAP parameters as in Ref.\ \citenum{Bartok2018}, to ensure comparability.

The task in fitting a GAP model is therefore to find the regression coefficients, $\alpha$. Assume we have a reference database of $n$ structures, for each of which we know DFT-computed energies, $E_{j}$, and forces, ${\vec F}_{j,j'}$, with scalar components in $x$, $y$, and $z$ direction for all $j'$ atoms in the $j$-th structure. (We also add DFT-computed virial stresses, which does not change the approach but makes the expressions more bulky, so we only discuss the case of energies and forces below.) We collect all entries of our DFT reference database in a single vector, ${\vec y}$:
\begin{equation}
    {\vec y} = \left( E_{1}, ..., E_{n}, F^{(x)}_{1,1}, ..., F^{(z)}_{n,j'} \right).
    \label{eq:y} 
\end{equation}
We then define another vector, ${\vec y'}$, which contains the atomic energies for all $N$ atoms in the reference database, which we {\em do not} know from DFT:
\begin{equation}
    {\vec y'} = \left( \varepsilon_{1}, ..., \varepsilon_{N} \right).
\end{equation}
We finally define a linear differential operator, ${\bf L}$, which connects the two:
\begin{equation}
    {\vec y} = {\bf L} {\vec y'}.
\end{equation}

The fitting coefficients are then obtained from the reference data, encoded by {\bf y}, according to \cite{QuinoneroCandela2005, Bartok2018}
\begin{equation}
    \alpha^{\ast} = \left[ {\vec K}_{MM} + ({\vec LK}_{NM})^{T} {\vec \Lambda}^{-1} {\vec LK}_{NM}
                    \right]^{-1}
                    ({\vec LK}_{NM})^{T} {\vec \Lambda}^{-1} {\vec y}
    \label{eq:alpha}
\end{equation}
with ${\vec K}$ denoting kernel matrices based on the $K$ defined above (here, SOAP), and the size of ${\vec K}_{NM}$ being the number of total atoms, $N$, times the number of sparse points, $M$. We emphasize that in computational practice, neither ${\vec L}$ nor ${\vec K}_{NM}$ are computed individually; only the combined matrix ${\vec LK}_{NM}$ is.

The entries of {\vec y} in the above equation ({\em i.e.}, our DFT reference data) were obtained using CASTEP 8.0, \cite{Clark2005} with on-the-fly pseudopotentials, the PW91 functional, \cite{Perdew1992} and a basis-set extrapolation scheme, \cite{Francis1990} and using the same convergence parameters as in GAP-18. \cite{Bartok2018} All new potential versions were fitted with the same parameters as GAP-18 (including the baseline pair potential, $V^{(2)}$; Eq. \ref{eq:E}), but varied in the composition of the reference database and the number of sparse points, $M$. All new potential versions will be deposited in a publicly available repository (Zenodo) upon publication of this work.

\subsection{Regularization}

The key aspect of Eq.\ \ref{eq:alpha}, in the context of the present work, is now the diagonal matrix ${\vec \Lambda}$, which contains the expected errors for all entries of $y$:
\begin{equation}
    {\vec \Lambda} =
    \begin{pmatrix}
     ( \sigma_{E}^{(1)} )^{2} & & & & & \\
                      & \ddots & & & & \\
                      & &  ( \sigma_{E}^{(n)} )^{2} & & & \\
                      & & & ( \sigma_{F}^{(1)} )^{2} & & \\
                      & & & & \ddots & \\
                      & & & & & ( \sigma_{F}^{(n')} )^{2} \\
    \end{pmatrix}
\end{equation}

In previous GAP fits, these expected errors have been set based on physical intuition and for separate parts of a given database (for example, assigning a smaller $\sigma_{E}$ to crystalline configurations and a larger $\sigma_{E}$ to amorphous ones). \cite{amoC, Bartok2018} Instead, we now use a protocol for fitting phonon properties based on simple supercell displacements and an adjusted regularization (expected error) in the fit; a similar idea has been used very recently in Ref. \citenum{Babaei2019}, but not yet explored in detail. We here set the atom-wise regularization for force components in $ {\vec F}_{i} = ( F^{(x)}_{i}, F^{(y)}_{i}, F^{(z)}_{i} ) $ according to 

\begin{equation}
    \sigma_{F}^{(i)} = 
    \begin{cases}
    f \times \left| {\vec F}_{i} \right| ,& {\rm if } \left| {\vec F}_{i} \right| >  F_{\rm min}; \\
    f \times F_{\rm min} ,& {\rm else.}
    \end{cases}
    \label{eq:sigma}
\end{equation}

Initially, we chose $f = 0.1$ and $F_{\rm min} =  0.01  \; {\rm eV \, \AA{}^{-1}}$. This value already leads to a lower bound of $\sigma_{\rm F} = 0.001  \; {\rm eV \, \AA{}^{-1}}$, compared with $\sigma_{\rm F} = 0.1 \; {\rm eV \, \AA{}^{-1}}$ for crystalline configurations in GAP-18. \cite{Bartok2018} We later vary $f$ over a wide range of values.

\subsection{Vibrational properties} 

Phonon and thermal property computations were carried out using phonopy \cite{Togo2015a} and phono3py \cite{Togo2015} with the finite displacement method. The size of the atomic displacements were 0.01 \AA{} (phonopy) and 0.03 \AA{} (phono3py), as per the standard setting. The thermal conductivity was calculated with the relaxation time approximation as implemented in phono3py. 

To compute second-order and third-order force constants for optimized structures with the help of GAP potentials, a script was built that uses pymatgen 2019.12.22,\cite{jain2013commentary} Atomic Simulation Environment (ASE) 3.19.0,\cite{larsen2017atomic} phonopy 2.4.2, phono3py 1.18.2 and quippy (including the GAP code, development version of 10 Jan 2020). The built-in geometry optimization of ASE was first used to optimize the crystal structures (including cell size) with the help of the GAP potentials. The forces on each atom were smaller than $10^{-5}$ eV/\AA{}. Phonon band structures were calculated along high-symmetry lines in reciprocal space; the latter were identified as described in Ref. \citenum{setyawan2010high} and as implemented in pymatgen. 51 points were calculated between each high-symmetry point. The supercells for the phonopy calculations were created based on the primitive cells and in such a way that each lattice parameter was larger than 15 \AA. To compare the DFT benchmark calculations of the harmonic phonons to GAP potentials, an RMS value was defined that compares all bands along each point in reciprocal space at which the bands were calculated. The supercells for the phono3py computations were based on the conventional cells and they were also built such that each lattice parameter was $> 15$ \AA. The thermal conductivity for diamond was calculated for the conventional cell and with an 11$\times$11$\times$11 $q$-point grid. The complete code will be made available {\em via} Github upon publication.

\begin{figure}[t]
\includegraphics[width=8cm]{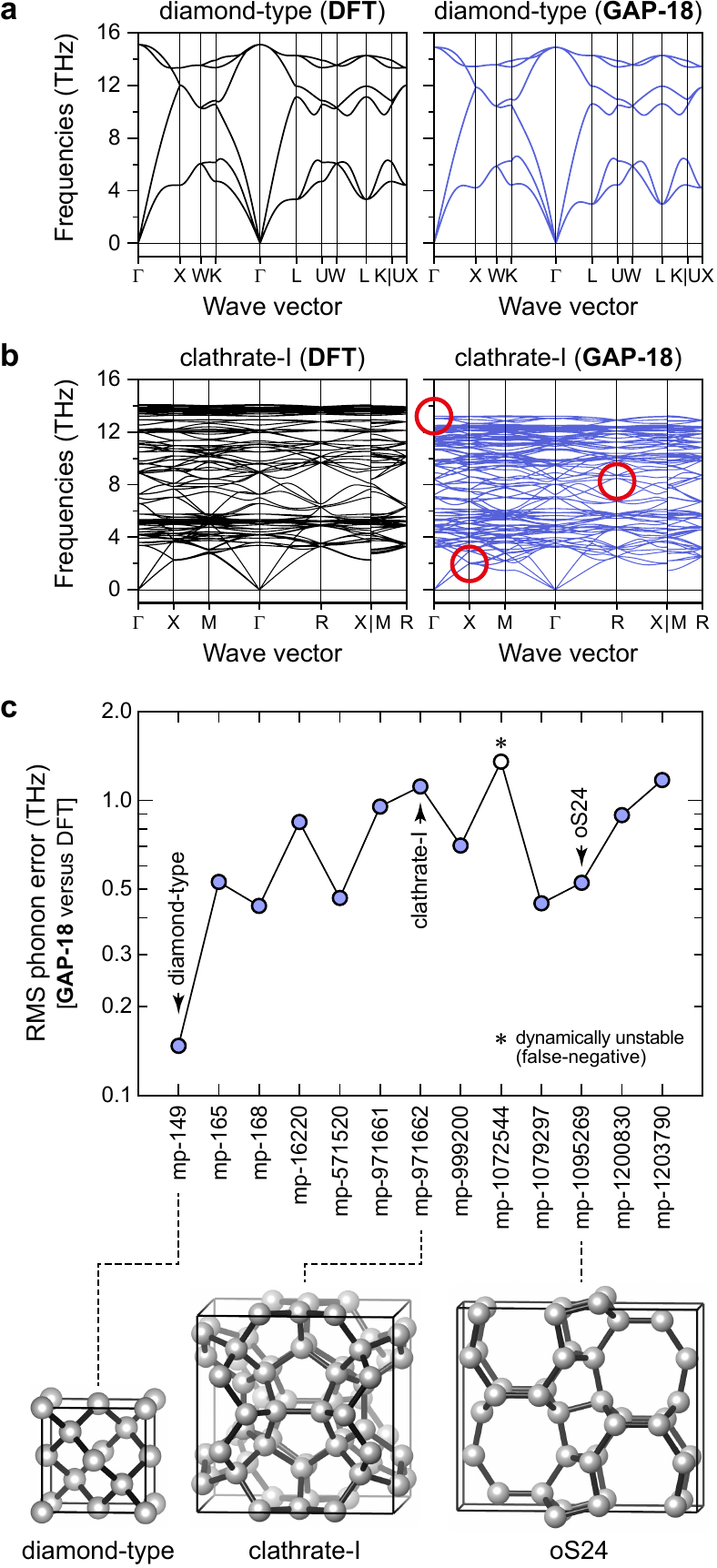}
\caption{The current state-of-the-art in predicting phonons with the GAP methodology, illustrated using the general-purpose GAP-18 model for silicon. \cite{Bartok2018} (a) Phonon dispersion curves for the {\bf dia}-type structure, computed with DFT ({\em left}) and GAP-18 ({\em right}), showing excellent agreement. (b) Same for clathrate-I-type silicon. The general features of the phonon dispersion are reproduced as well, but there are several discrepancies in detail (highlighted in red), which are discussed in the text. (c) A survey of all ambient-pressure stable structures from the Materials Project (using the ``mp'' identifier) up to a reasonable size, and the root-mean-square error for their respective GAP-18 predicted phonon dispersions. Lines connecting data points are guides to the eye.}
\label{fig:gap-18-intro}
\end{figure}

\cleardoublepage

\begin{figure*}[t]
\includegraphics[width=17cm]{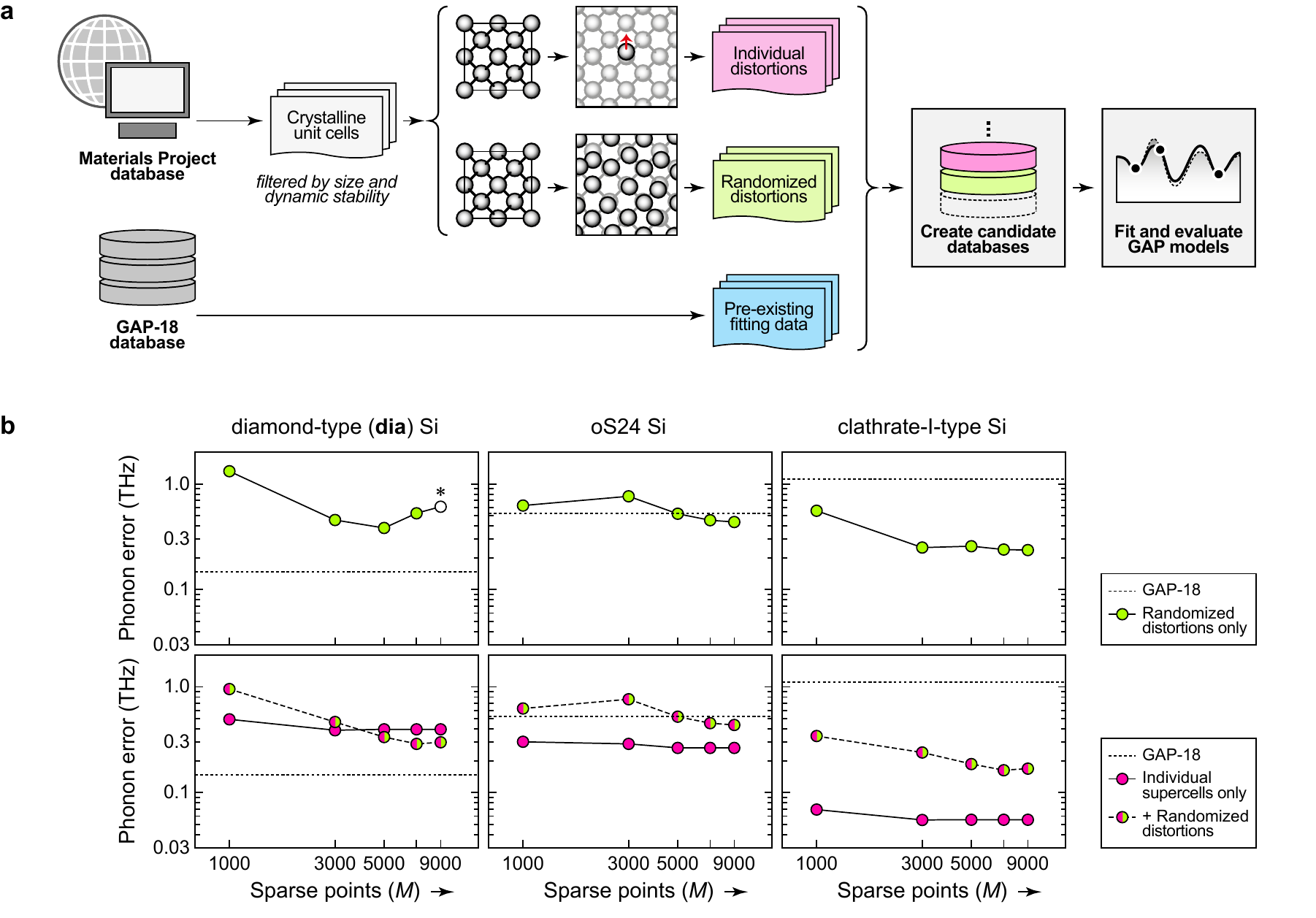}
\caption{(a) Strategies for building reference databases, shown by a simplified schematic. We extract crystal structures from the Materials Project database, with filtering as described in the text. We then create supercells based on these structures in which we either displace atoms individually, or all of them randomly, and compute DFT energies and forces for these supercells. Additionally, we investigate the effect of including the pre-existing GAP-18 database. \cite{Bartok2018} We then build various combinations of databases to which candidate GAP models are fitted. (b) Results for various database building schemes sketched above, shown as phonon-error ``learning curves'' for three representative Si allotropes. An asterisk ($\ast$) indicates a structure that is erroneously predicted as dynamically unstable. The reference values for GAP-18 are shown by dotted horizontal lines. }
\label{fig:overview}
\end{figure*}

\section{Results and discussion}

\subsection{State of the art}

We begin by discussing the performance of GAP-18, \cite{Bartok2018} a general-purpose ML potential. We quantify how well it can predict the phonon dispersion relations for diamond-type ({\bf dia}) silicon, which is abundantly represented in its reference database (489 pristine and 404 defective bulk cells, some containing $> 200$ atoms per cell, plus various surface configurations; Ref.\ \citenum{Bartok2018})---but also for other allotropes which GAP-18 has not ``seen''. We compute all reference phonon band structures at the same level of DFT, enabling direct benchmarking. In the present work, we consider only structures without external pressure, but we mention that high-pressure silicon allotropes have been successfully studied with ML potentials before. \cite{Behler2008, Behler2008a}

For {\bf dia}-Si (Fig.\ \ref{fig:gap-18-intro}a), GAP-18 predicts phonons with practically quantitative accuracy. \cite{Bartok2018} At this level of quality, the computed phonon band structures from both methods are practically indistinguishable to the naked eye. We determined the root-mean-square (RMS) error of phonon eigenvalues across the Brillouin zone, which amounted to 0.15 THz. 

The situation is different for the clathrate-I-type structure which is not included in the GAP-18 reference database (Fig.\ \ref{fig:gap-18-intro}b). Here, using the potential out-of-the-box still leads to a dynamically stable structure ({\em i.e.}, without any imaginary eigenvalues, which would conventionally be plotted as negative frequencies), and it does recover the general features. There are, however, notable quantitative differences, and some of them are highlighted in Fig.\ \ref{fig:gap-18-intro}b by red circles. The highest-energy phonon mode at the zone center, $\Gamma$, is under-predicted by GAP by 0.8 THz (6\%) whereas the lowest predicted band at the $X$ point deviates by 0.3 THz (12\%) from the DFT reference data. Especially the latter relative error is much larger (more than twice as large) than relative errors typically arising from DFT computations in comparison to experimental measurements: The mean relative error from DFT to experiment calculated from 53 materials was --3.6\% in a recent high-throughput study.\cite{petretto2018high} Errors arising from ML potentials should therefore be in a similar range or smaller. The red circle at the $R$ point further highlights that the phonon frequencies calculated by GAP-18 are smaller than the ones calculated by DFT. The RMS error is 1.11 THz, almost ten times that obtained for {\bf dia}-Si.

To obtain a more comprehensive picture, we test the root-mean-square (RMS) error of phonon eigenvalues for a wide range of dynamically stable structures from the Materials Project database \cite{jain2013commentary} (Fig.\ \ref{fig:gap-18-intro}c). Out of 13 allotropes considered, all except one are correctly predicted to be dynamically stable. The exception is mp-1072544, a hypothetical structure. \cite{Zhao2013} The phonon prediction error for most allotropes is about 0.5--1.0 THz, with clathrate-I being one of the more poorly described examples. This is still a remarkable quality for structures which have not been included in the GAP-18 reference database, such as clathrate-I or oS24. \cite{Kim2015} But the error is clearly too large for {\em quantitative} studies or for a fully reliable assessment of dynamic stability, which one might wish to carry out when computationally screening  large amounts of hypothetical structures.

\subsection{Approaches for building fitting databases}

We now perform a comprehensive study of how reference databases can be designed for fitting phonon-accurate GAPs if {\em no} prior such database exists. Figure \ref{fig:overview}a provides an overview. We extract structures from the Materials Project database and filter them according to a maximum system size and the criterion that the structure has not significantly changed after the initial structural optimization with CASTEP. This structural change was determined by the StructureMatcher routine implemented in pymatgen, using slighly tighter tolerances for matching structures (ltol = 0.1, stol=0.1 and angle\_tol=3) instead of the default parameters.\cite{jain2013commentary} We then create supercell models with atomic displacements and generate DFT reference data for those, using two different strategies. A common approach uses randomly displaced cells, creating several copies with lattice parameters scaled by a few percent, and atomic positions randomized with a standard deviation of 0.01 \AA{}, for example (green in Fig. \ref{fig:overview}a). We also test a different approach (magenta): create supercells of the optimized crystal structures with (only) individual displacements and atom-wise regularization, as discussed in Sec.\ \ref{sec:methods}. We finally fit candidate GAPs to the resulting databases, always using the same descriptors as in GAP-18, and evaluate their phonon errors by the same RMS measure as in Fig.\ \ref{fig:gap-18-intro}. 

The randomized structures, as used in previous GAPs ({\em e.g.}, Ref.\ \citenum{amoC}), provide an acceptable but not a truly reliable result (Fig.\ \ref{fig:overview}b, top row). We study the evolution of the error as a function of the number of sparse points, $M$, used in the fit. The outcome of this procedure varies notably from structure to structure; the prediction quality is better than GAP-18 for clathrate-I, about on par for the open-framework oS24 allotrope, but {\em worse} than GAP-18 for diamond-type silicon. This is indicative of the fact that a similar (and rather small) number of 10 distorted supercells is used to represent each structure, with a view to keep the computational workload tractable even for more complex chemical systems. The number of {\bf dia}-like atomic environments in our databases is therefore much smaller than in that of GAP-18. Interestingly, the {\bf dia} ``learning curve'' does not converge with $M$ as expected, instead leading to a false-positive prediction of dynamic instability (``$\ast$'' in Fig. \ref{fig:overview}b). 

An alternative strategy is to create individual displacements (only), in separate supercells, akin to the way that one would build supercells for phonon computations. Displacements are generated along all symmetry-inequivalent directions, which leads to only one supercell for several structures including {\bf dia} ($Fd\bar{3}m$), but many supercells for other structures, depending on space-group symmetry. In this case (magenta in Fig.\ \ref{fig:overview}), the phonon accuracy is generally better than with randomized distortions only, and it plateaus after about 3,000 representative atoms---this is clearly expected because the total number of {\em relevant} force-component entries in ${\vec y}$ (Eq.\ \ref{eq:y}) is relatively small, mostly relating to the displaced atom in a supercell and those atoms in its local environment. There are, again, pronounced differences for the three allotropes, with clathrate-I again being most accurately described among those characterized in Fig.\ \ref{fig:overview}b. 

We also test combined models, which include both Materials Project derived databases (and only those), indicated in Fig.\ \ref{fig:overview}b by mixed magenta/green symbols. This appears to even out the performance for the three different allotropes, increasing the phonon error for clathrate-I-type silicon, albeit not above 0.2 THz. The learning curves for the combined databases converge more slowly with $M$ than those for only individual displacements.

\begin{figure}[t]
\includegraphics[width=8cm]{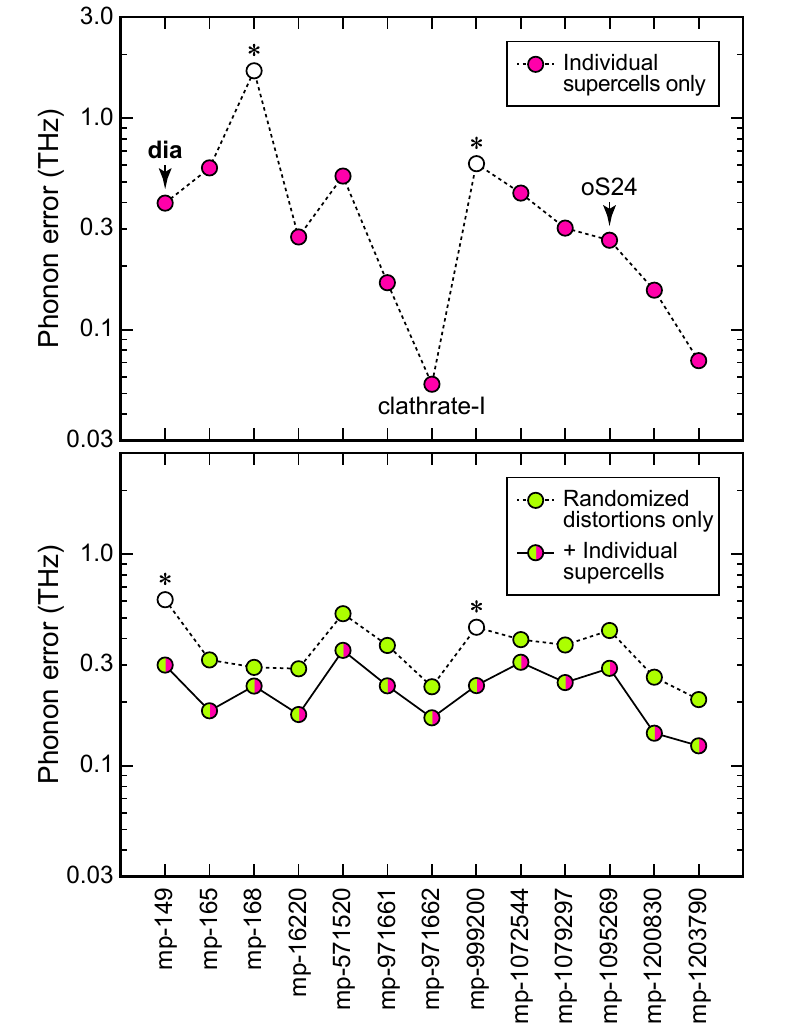}
\caption{Predicting phonon frequencies ``from scratch'' based on Materials Project entries. RMS phonon band structure errors are given for all relevant MP entries as in Fig.\ \ref{fig:gap-18-intro}c, but now using new potentials, fitted to reference databases created as outlined in Fig.\ \ref{fig:overview}a. All potentials use $M = 9000$ sparse points. {\em Top:} Results from the fit with only individual supercells. Two structures are erroneously predicted to be dynamically unstable ($\ast$). {\em Bottom:} Same for a potential fitted to randomly distorted structures only (green) and to a combined database (green/magenta).} 
\label{fig:compare_allMP}
\end{figure}

Having assessed the qualitative and individual performance of the method for selected structures, we now re-visit the full range of relevant silicon allotropes, with results for RMS phonon errors collected in Fig.\ \ref{fig:compare_allMP}. To make the comparison easier, we always choose $M = 9000$, as in GAP-18.

Mirroring now more broadly what was already observed in Fig.\ \ref{fig:overview}b, the prediction quality is overall quite scattered if only invididual supercells are used to construct the reference database (top panel). For most of the silicon allotropes, the approach leads to a prediction error of better than 0.5 THz; clathrate-I turns out to be the best described of all of them. When randomized distortions are used, on the other hand, the distribution of prediction errors is more uniform (green points in the bottom panel in Fig.\ \ref{fig:compare_allMP}). Two of the 13 structures are erroneously predicted to be dynamically unstable with both approaches---one of them being {\bf dia}-Si in the case of using only randomized distortions, as noted above.

Adding individual supercells to the randomized distortions, {\em i.e.} combining both strategies outlined at the top of Fig.\ \ref{fig:overview}a, appears to improve the results throughout. This is an important general finding: our individual-supercell strategy (magenta) can, and apparently should, be combined with other ways of sampling configuration space when developing GAP fitting databases. We emphasize that this protocol can be fully automated, requiring only a choice of the magnitude of the displacements and of the number of supercells to be created per structure; it is therefore expected to be easily coupled to existing and new high-throughput databases of crystal structures. An open research question, beyond the scope of the present work, is how these automated potentials can accommodate previously unseen structures which are not drawn from the Materials Project or another database, but are discovered, for example, during a GAP-driven random search. \cite{GAP-RSS_npj}

\begin{figure}[t]
\includegraphics[width=8.0cm]{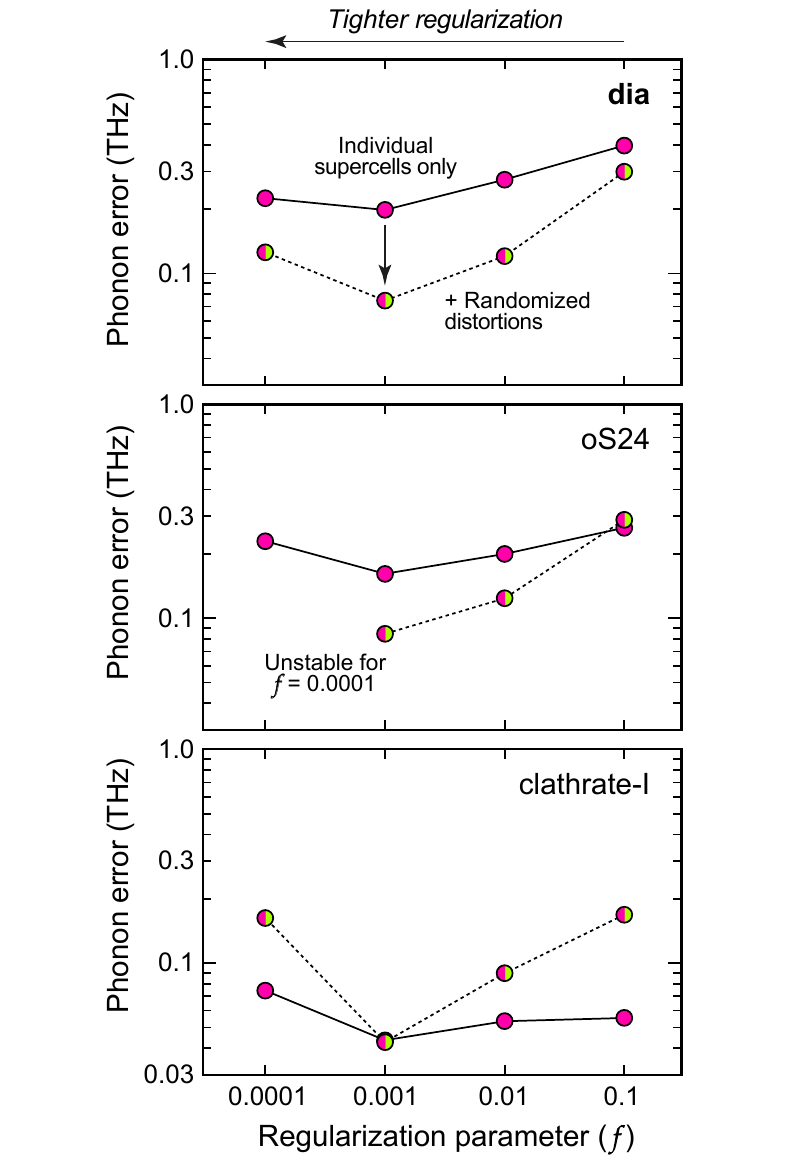}
\caption{Optimization of the regularization parameter $f$, as defined in Eq.\ \ref{eq:sigma}. We performed series of fits, either with only the individually distorted Materials Project based supercells (magenta) or with the random displacements added (magenta / green). RMS errors for computed vibrational eigenvalues are given as before. The optimal value among those we tested, $f = 0.001$, is indicated by arrows. For oS24, the combined fit at $f = 0.0001$ leads to a change in structure during optimization, as a result of overfitting, and therefore no RMS error for the phonon band structure can be obtained. This indicates the reasonable limit within which $f$ should be chosen.}
\label{fig:regularisation}
\end{figure}

\subsection{Optimized regularization}

In this section, we analyze in more detail the effect of the atom-wise regularization on the GAP prediction of phonon band structures. This corresponds to the idea of making the fit ``looser'' (more flexible) or ``tighter'' (more accurate in the required regions of configuration space, but therefore less flexible), controlled by the single factor $f$ which we use to scale all atom-wise force regularization parameters according to the absolute force on that atom (Eq. \ref{eq:sigma}). We now perform GAP fits to versions of the ``magenta'' part of the database (Fig. \ref{fig:overview}a) in which we vary $f$ over several orders of magnitude, and inspect its effect on the performance for {\bf dia}, oS24, and clathrate-I-type silicon. We also test whether the additional inclusion of randomized distortions in the reference database would still improve the fit (as suggested by Fig. \ref{fig:compare_allMP}). We remind the reader that we had so far used $f=0.1$ throughout, and we now tighten the regularization, down to $f=0.0001$. These results are collected in Fig. \ref{fig:regularisation}.

For {\bf dia} silicon, the phonon prediction error initially improves substantially when lowering $f$, {\em viz.} from about 0.4 to about 0.2 THz. However, for $f = 0.0001$, the error increases again, indicating a too low ``expected error'' for the input data. The quality of the fit can be even further improved, to better than 0.1 THz, if randomized distortions are added to the fitting database (note that we do not use atom-wise regularization for those configurations, instead employing the GAP-18 default of $\sigma_{F}=0.1$ eV \AA{}$^{-1}$, to retain higher flexibility for those more disordered configurations). For oS24, a qualitatively similar trend is observed in Fig. \ref{fig:regularisation}, although here the effect of too small $f$ is more drastic: the resulting potential led to a different structure during the relaxation that must be performed prior to evaluating the phonons, and therefore no RMS error for the latter can be obtained. This example more generally emphasizes the need for a (reasonably large) regularization of the GAP fit to arrive at stable results in practice. For clathrate-I-type silicon, the data shown in Fig. \ref{fig:regularisation} similarly suggest the existence of an optimal $f$, and here we observe an overall increase in phonon error when randomized configurations are added to the individually displaced supercells---a consequence of the ``equalization'' between different structures that had already been observed in the previous subsection. Taken together, these results suggest a choice of $f$ between 0.01 and 0.001, but not smaller.

We note that ongoing work in the community aims to optimize GAP fits with regard to the basis functions and hyperparameters for the structural descriptor. \cite{Willatt2018, Caro2019, Himanen2020} The present study complements these efforts by probing the question how the regression task itself might be further optimized. 

\subsection{Transferability: Extending GAP-18}

\begin{figure}[t]
\includegraphics[width=8.0cm]{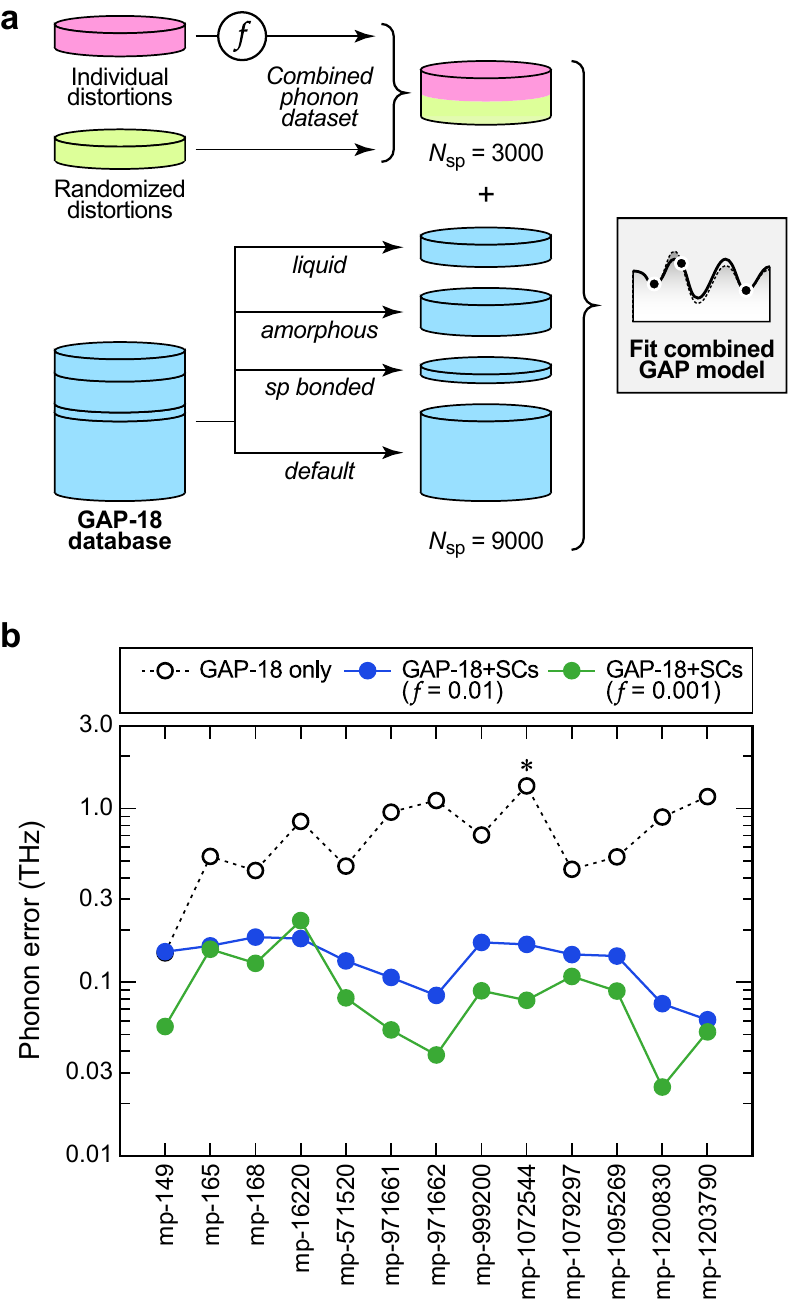}
\caption{
(a) Building a reference database for the combined potential. Separate configuration type labels are set for the different parts of the GAP-18 database (blue), which are detailed in Ref.\ \citenum{Bartok2018}, and a new type is introduced for the datasets which we add in the present work (magenta/green). The number of sparse points, $N_{\rm sp}$, for each configuration type is given; their sum is the total number of sparse points, $M = 12000$.
(b) Prediction quality for phonons across the range of relevant structures -- shown as in Fig. \ref{fig:compare_allMP}, but now comparing GAP-18 and extended potentials that are based on it. In both of these, the individually displaced and randomized supercells (``SCs'') have been added to the fit, with atom-wise regularization according to $f = 0.01$ and $f = 0.001$ (cf.\ Eq.\ \ref{eq:sigma}), respectively. All structures are correctly predicted to be dynamically stable with both of the modified potentials. }
\label{fig:MP_structures}
\end{figure}

We now take our newly created reference databases, constructed from individual distortions with adaptive regularization (magenta in Fig.\ \ref{fig:overview}a) and randomized supercells (green in Fig.\ \ref{fig:overview}a), respectively, and combine them with the existing database to which the GAP-18 model had previously been fitted \cite{Bartok2018} (blue in Fig.\ \ref{fig:overview}a). We thereby aim to answer another more general question about GAPs---namely, how can existing potentials be extended? And how will these potentials retain their previously validated properties, {\em viz.} for GAP-18, the all-round accurate description of {\bf dia}, liquid, and amorphous silicon?

Our combined fits use a total of $M=12000$ sparse points, of which 9000 are drawn from the previous database, with distribution among different configuration types (liquid, amorphous, {\em etc.}) as in GAP-18, \cite{Bartok2018} and 3000 are drawn from a combined set of individually and randomly displaced supercells (Fig.\ \ref{fig:MP_structures}a). We combine both new databases because Fig.\ \ref{fig:compare_allMP} had indicated that doing this may lead to a more robust behaviour than using the individually displaced cells on their own. Based on our analysis of the regularization parameter (Fig.\ \ref{fig:regularisation}), we generate two candidate potentials with $f=0.01$ or $f=0.001$, respectively.

Figure \ref{fig:MP_structures}b shows that both potentials reach a prediction quality within at least the 0.1--0.2 THz range, with even better performance for some structures. There is a further minor improvement when moving from $f=0.01$ to $f=0.001$, exemplified by an improved description of the {\bf dia} structure with smaller $f$, in line with what the tests in Fig.\ \ref{fig:regularisation} have shown. But both potentials would clearly seem useful for practical purposes when judged on the quality of their phonon predictions alone.

\begin{table}[t]
\caption{
Thermal conductivity, $\kappa$, for {\bf dia}-Si as obtained from GAP-18 and extended versions of it, compared to DFT and experimental references at different temperatures.}
 \begin{tabular}{l c c c} 
 \hline
 \hline
  & \multicolumn{3}{c}{$\kappa$ (W m$^{-1}$ K$^{-1}$)} \\
 \cline{2-4}
 Method & 300 K & 600 K & 900 K \\ [0.5ex] 
 \hline
 Expt. (Ref. \citenum{shanks1963})&142.2&69.2&33.7\\
 \hline 
 DFT (PBE, Ref. \citenum{Ouyang2017}) &137.4 & & \\
 DFT (PW91, this work) & 126.4  &57.1  & 37.4 \\ 
 \hline 
 GAP-18 & 109.4 &50.2& 32.9    \\
 GAP-18+SCs ($f=0.01$)&  109.8  &  50.3 & 33.0 \\ 
 GAP-18+SCs ($f=0.001$) & 53.7  &  26.5 & 17.7 \\
 \hline
 \hline
\end{tabular}
\label{tab:thermalcond}
\end{table}

To test transferability of the newly created potentials, we compared the calculated thermal conductivity for {\bf dia} with DFT references and experimental data. This is a property which is reasonably well described by GAP-18 and the question is therefore whether our modifications lead to a detriment in that performance. To calculate accurate thermal conductivity, third-order force constants have to be calculated. Here, this is done with the help of a finite displacement method where a pair of atoms within the supercell has to be displaced. Such cells are not part of our reference data, and therefore the potential needs to ``pick up'' the corresponding (anharmonic) physics from other cells in the database, such as MD snapshots.  The previous version of GAP-18 contains plenty of the latter, and therefore it arrives at acceptable values for the thermal conductivity in comparison to our DFT reference values (less than 15\% deviation at each temperature, Tab. \ref{tab:thermalcond}). We mention in passing the good performance of independent, more specialized GAP and neural-network potential models for predicting $\kappa$. \cite{Babaei2019, Minamitani2019, Qian2019} 

GAP-18+SCs ($f=0.01$) shows the same results as the original GAP-18 within any reasonable accuracy, whereas $f=0.001$ leads to an entirely unreliable prediction of $\kappa$. This is now a crucial point, because it indicates that the latter fit is too ``tight'' on the newly added structures (which correspond to harmonic phonons only), and therefore leads to overfitting behavior which manifests in poor prediction of forces in other regions of configuration space. It also emphasizes that the prediction quality for $\kappa$ is determined by the GAP-18 dataset. Follow-up work on optimized databases with regard to predicting thermal conductivity and related properties is currently being planned.

\begin{figure}[t]
\includegraphics[width=8.0cm]{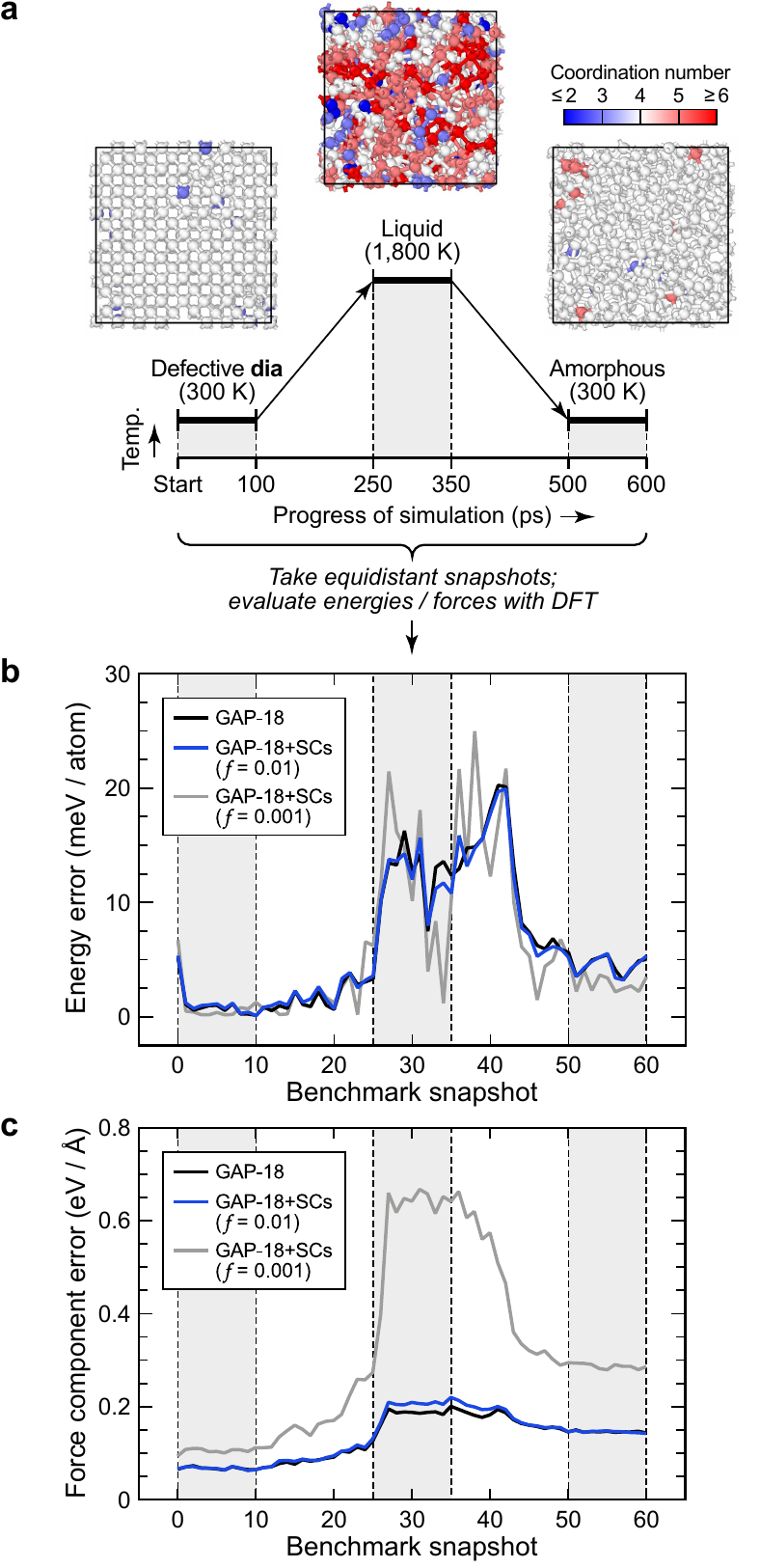}
\caption{Assessing the transferability of extended potentials by using a systematic benchmark for disordered structures. (a) Overview of the protocol: a defective diamond-type structure (500 atoms) is thermalized in constant-pressure molecular-dynamics simulations driven by GAP-18, heated to 1,800 K and cooled again (at 10$^{13}$ K s$^{-1}$) to form a quenched amorphous state, similar to Ref.\ \citenum{aSi_structures_GAP}. From this trajectory, we take structural snapshots every 10 ps (10,000 simulation steps), and evaluate their energies and forces with DFT. (b) Prediction error for the per-atom energy from various GAP models, given as absolute error versus DFT. (c) Same but now for the forces, with errors given as RMS over all Cartesian force components in a given structure.}
\label{fig:mq}
\end{figure}

\cleardoublepage

\begin{table}[t]
\caption{
Numerical errors for the melt--quench test, as illustrated in Fig.\ \ref{fig:mq}a, as a concise quality measure for the different potentials. Errors are averaged over the last five snapshots for each respective part of the trajectory. }
 \begin{tabular}{l c c c} 
 \hline
 \hline
  & Defective {\bf dia} & Liquid & Amorphous \\
  & (300 K) & (1,800 K) & (300 K) \\ [0.5ex] 
 \hline
 \multicolumn{4}{l}{{\em Energy error (meV / atom)}} \\ 
 \hline 
 GAP-18 & 0.5	&	12.2	&	4.3 \\
 \quad +SCs ($f=0.01$)  &  0.6	&	11.5	&	4.2 \\ 
 \quad +SCs ($f=0.001$) &  0.6	&	8.4	&	2.6 \\
 \hline
 \multicolumn{4}{l}{{\em Force component error (eV / \AA{})}} \\ 
 \hline 
 GAP-18 & 0.07	&	0.19	&	0.15 \\
 \quad +SCs ($f=0.01$)  &  0.07	&	0.21	&	0.14 \\ 
 \quad +SCs ($f=0.001$) &  0.11	&	0.65	&	0.29 \\ 
 \hline
 \hline
\end{tabular}
\label{tab:mq}
\end{table}

We finally address the question whether the new potentials can also be used in their initial design space, for example, for disordered phases. GAP-18 has enabled accurate atomistic studies of liquid and amorphous silicon \cite{aSi_structures_GAP, Bernstein2019} and we therefore investigate whether the extended versions are still applicable to the same problems. We designed a test that, for the first time to our knowledge, critically assesses the energy and force accuracy during GAP-driven melt--quench simulations. We do this by running an example trajectory that is representative of ``real-world'' applications, collecting an ensemble of structures, and benchmarking against DFT data.

The starting point (Fig.\ \ref{fig:mq}a) is a 512-atom supercell of diamond-type silicon in which 12 atoms are randomly removed, leading to defects including a vacancy cluster. We run a GAP-MD simulation over several hundreds of thousands of steps, in the NPT ensemble as implemented in LAMMPS, \cite{Plimpton1995} similar to Ref.\ \citenum{aSi_structures_GAP}. At every 10 ps, we take a snapshot and compute its energies and forces using CASTEP \cite{Clark2005} at the PW91 level. \cite{Perdew1992} 

The energy error (Fig.\ \ref{fig:mq}b) is instructive to watch for the various new GAPs. It traces the GAP-18 result almost perfectly for $f=0.01$, but fluctuates much more strongly with $f=0.001$, already indicating that this model might not completely recover the behavior of the initial potential on which it is based. However, the {\em absolute} energy error is superficially not bad, and this has an important message for the benchmarking of GAPs: the average energy error itself might not provide sufficient information. Indeed, the averaged energy error appears to be {\em best} for $f=0.001$ in the final part of the liquid trajectory, {\em viz.} 8 meV per atom compared to 12 for both other potentials (Table \ref{tab:mq}). A strikingly different result, however, is seen in the force components (Fig.\ \ref{fig:mq}c), where $f=0.01$ recovers the behavior of GAP-18 whereas $f=0.001$ leads to more than triple the error. In summary, our extended potential with $f=0.01$ does not lead to any notable loss in quality for disordered silicon (Table \ref{tab:mq}) whilst improving the description of phonons by almost an order of magnitude (Fig.\ \ref{fig:MP_structures}).

\section{Conclusions}

The ability to compute accurate vibrational and thermal properties with ML potentials is an important result for two reasons. First, it has provided a useful testing ground for the way that reference databases for such potentials can be constructed with minimal computational effort---the main focus here being on the ability to treat various crystal structures {\em at the same time}. This may lead to a better understanding of the role and physical meaning of the ``expected error'' built into the GAP fit. Second, it is a step forward towards automated computational materials science, enabling the possibility of high-throughput vibrational property prediction especially when coupled to efficient workflows. \cite{Ong2013, Pizzi2016, mathew2017atomate} 

While most computational work for silicon has focused on the diamond-type allotrope, it is expected that the search for other structures and thereby synthesis targets might be productive. The theory-guided discovery and, ultimately, synthesis of new materials has now been achieved in many cases \cite{Oganov2019} and it might be accelerated by ML-driven phonon computations as presented here. For the specific example of group-14 elements, we mention the successful synthesis of clathrate-II-type germanium \cite{Guloy2006} and also the wealth of intercalated clathrate compounds, \cite{Beekman2008} which may now be addressed with similar computational methodology.

\section*{Acknowledgements}
J.G. acknowledges support within the HPC-Europa3 programme (INFRAIA-2016-1-730897), with the support of the EC Research Innovation Action under the H2020 Programme, and from the European Union’s Horizon 2020 research and innovation program under the Marie Sk\l{}odowska-Curie grant agreement No 837910. V.L.D. acknowledges a Leverhulme Early Career Fellowship. We are grateful for computational support from the UK national high performance computing service, ARCHER, for which access was obtained via the UKCP consortium and funded by EPSRC grant ref EP/P022561/1. We also acknowledge the Consortium des  \'E{}quipements de Calcul Intensif en F\'e{}d\'e{}ration Wallonie Bruxelles  (C\'E{}CI) for computational resources. 
Structural drawings were created using VESTA \cite{Momma2011} and OVITO.\cite{Stukowski2010}
We thank Atsushi Togo for helpful input.

\section*{Conflicts of interest}
A.P.B. and G.C. are listed as inventors on a patent filed by Cambridge Enterprise
Ltd. related to SOAP and GAP (US patent 8843509, filed on 5 June
2009 and published on 23 September 2014). The other authors declare
no conflict of interest.

\section*{Author contributions}
J.G. and V.L.D. initiated and coordinated the project. J.G. designed and implemented the workflows for phonon and thermal property analyses. G.C. and A.P.B. proposed the force-proportional atom-wise regularisation; A.P.B. implemented it in code; G.C. provided advice on fitting. V.L.D. developed the reference databases and fitted the modified potentials. J.G. and V.L.D. wrote the paper with input from all authors.

\section*{Data availability}
The data that support the findings of this study will be made openly available in a Zenodo repository after publication.

\end{document}